\documentclass[aps,prb,preprint,groupedaddress,showpacs]{revtex4}

\usepackage{amsmath}
\usepackage{amssymb}
\usepackage{epsfig}
\newcommand {\dr}{{\mathrm d}\mathbf{r}}

\newcommand {\dk}{{\mathrm d}\mathbf{k}}
\newcommand {\dd}{{\mathrm d}}
\newcommand {\rr}{\mathbf{r}}
\newcommand {\kk}{\mathbf{k}}
\newcommand {\etal}{\begin{itshape}et al\end{itshape}.}

\begin{document}


\title{Dynamical density functional theory and its application to spinodal
decomposition}


\author{A.J. Archer}
\email[]{Andrew.Archer@bristol.ac.uk}
\author{R. Evans}
\affiliation{H.H. Wills Physics Laboratory,
University of Bristol, Bristol BS8 1TL, UK}


\date{\today}

\begin{abstract}
We present an alternative derivation of the dynamical density functional theory
for the one body density profile of a classical fluid developed by Marconi and
Tarazona [{\it J.\ Chem.\ Phys.}, {\bf 110}, 8032 (1999)]. Our derivation
elucidates further some of the physical assumptions inherent in the theory
and shows that it is not restricted to fluids composed of particles interacting
solely via pair
potentials; rather it applies to general, multi-body interactions. The
starting point for our derivation is the Smoluchowski equation and the theory
is therefore one for Brownian particles and as such is applicable to
colloidal fluids. In the second part of this paper we use the dynamical density
functional theory to derive a theory for spinodal decomposition that is
applicable at both
early and intermediate times. For early stages of spinodal decomposition
our non-linear theory is equivalent to the (generalised) linear
Cahn-Hilliard theory, but for later times it incorporates coupling
between different Fourier components of the density fluctuations (modes) and
therefore goes beyond Cahn-Hilliard theory. We describe the results of
calculations for a model (Yukawa) fluid which show that the coupling leads to
the growth of a second maximum in the density fluctuations, at a wavenumber
larger than that of the main peak.
\end{abstract}


\maketitle

\section{Introduction}
Classical density functional theory (DFT) \cite{Evans92} is a remarkably
successful theory for describing the rich behaviour of the equilibrium
structure and thermodynamics of fluids in external potentials. DFT has been used
to describe a wide variety of fluid interfacial, confinement and even freezing
phenomena; for example, DFT has
been a vital tool in understanding the wetting behaviour and surface phase
transitions of fluids adsorbed on
various substrates \cite{Evans92,Dietrich}. Given the success of DFT for
describing static inhomogeneous fluids, it is of great interest to be
able to build upon and incorporate these theories into a theory for the
{\em dynamics} of inhomogeneous fluids.

There have been several approaches to obtaining an equation of motion for the
one body density profile $\rho(\rr,t)$ of a classical fluid. The form that these
theories takes depends somewhat on how the particular
authors define $\rho(\rr,t)$. For some,
$\rho(\rr,t)$ is an ``ensemble'' average over the possible
configurations of the system at time $t$, given an ensemble of starting
configurations at an earlier time $t=0$. Using
this definition for $\rho(\rr,t)$,
there is clearly a unique density profile $\rho(\rr,t)$ at a given time $t$, and
therefore the equation governing the dynamics of this density profile will be
deterministic. This is the philosophy behind the approach of Marconi and
Tarazona (MT) in Refs.\ \cite{Marconi:TarazonaJCP1999,Marconi:TarazonaJPCM2000}.
Their approach obtains more formally some of the results proposed by earlier
authors, such as Evans \cite{Evans79} and Dieterich \etal\ \cite{Dieterich},
where it is assumed from the outset
that the gradient of the chemical potential $\nabla \mu(\rr,t)$ is the
thermodynamic force driving a particle current
\begin{equation}
\mathbf{j}(\rr,t)=-\Gamma \rho(\rr,t) \nabla \mu(\rr,t),
\label{eq:current}
\end{equation}
where $\Gamma$ is a mobility constant. An expression for $\mu$ is obtained
within DFT by assuming that, as in the case of the equilibrium fluid, the
chemical potential is given by the functional derivative of the Helmholtz free
energy functional
with respect to the density profile \cite{Evans79,Dieterich}. This assumption,
together with Eq.\ (\ref{eq:current}) and the continuity equation
\begin{equation}
\frac{\partial \rho(\rr,t)}{\partial t}\,=\,-\nabla . \mathbf{j}(\rr,t),
\label{eq:cont_eq}
\end{equation}
provide a basis for the deterministic formulation of dynamical DFT (DDFT).
Equations of this form have been
used, for example, to study the dynamics of freezing \cite{BagchiPhysicaA1987}
and of solvation \cite{Yoshimori:Day:Patay:JCP1998}. More recently, the more
systematic approach of MT has been used with much success to describe
the dynamics for several different systems. These applications refer to
particles in various external, time dependent, potentials
\cite{joe:christos,Flor1,Flor2,joepreprint}. For the systems considered,
the agreement between theory and the results from Brownian dynamics simulations
have generally been very good.

An alternative approach to obtaining a DDFT
is to view the fluid one body density profile (denoted
$\bar{\rho}(\rr,t)$ in order to distinguish it from $\rho(\rr,t)$, the
``ensemble'' averaged density profile) as some sort of spatial and/or time
coarse grained average of the density operator $\hat{\rho}(\rr,t)=\sum_{i=1}^N
\delta (\rr-\rr_i(t))$, where the $\rr_i(t)$ are the positions of the $N$
particles in the system. In this case, for Brownian particles, the equation
governing the dynamics of $\bar{\rho}(\rr,t)$ will, of
course, still contain a stochastic element. This is the viewpoint of a number of
theories \cite{KawasakiPhysicaA1994,
Kirkpatrick:ThirumalaiJPhysA1989, Munakata1, Munakata2, 
Munakata3, DeanJPhysA1996}. There is some confusion in the literature
\cite{DeanJPhysA1996,Frusawa:HayakawaJPhysA2000}
(see also Ref.\ \cite{KawasakiJPCM2000}) as to what precisely is meant by
$\rho(\rr,t)$. We will attempt to clarify these issues elsewhere
\cite{Rauscher}.

In this paper we adopt the deterministic viewpoint of MT and
consider $\rho(\rr,t)$ to be an ensemble average, i.e.\ an average over all
realisations of the stochastic noise in the preceding interval, until time
$t$. We employ the Smoluchowski equation as the starting point for an
alternative derivation of the DDFT for classical fluids developed by
MT in Refs.\
\cite{Marconi:TarazonaJCP1999,Marconi:TarazonaJPCM2000}. The present scheme for
deriving the DDFT bears some similarity in spirit to that given recently
in Ref.\ \cite{MunakataPRE2003}, where projector-operator techniques are used to
obtain first the Smoluchowski equation and subsequently an equation of motion
for the
fluid one body density profile -- the DDFT. Before proceeding with our
derivation of the DDFT, in Sec.\ \ref{sec:1} we give a brief introduction to the
Smoluchowski equation, expounding some of the physical assumptions concerning
the dynamics of the fluid that are implicit in this equation. In Sec.\
\ref{sec:2} we proceed with the derivation of the DDFT of MT from the
Smoluchowski equation. Sec.\ \ref{sec:3} describes an important application of
the DDFT to analyse the short and intermediate-time
dynamics of spinodal decomposition relevant to colloidal fluids. Finally in
Sec.\ \ref{sec:conc} we make some concluding remarks.

\section{The Smoluchowski equation}
\label{sec:1}
The Smoluchowski
equation \cite{Pusey,PuseyTough,Croxton,Risken,Gardiner} is a Fokker-Planck
equation (or
generalised diffusion equation) for interacting Brownian particles. A physically
intuitive way of arriving at this
equation, e.g.\ Ref.\ \cite{PuseyTough}, proceeds as follows:
For a fluid of $N$ Brownian particles, one imagines applying a force on the
particles, where the force on the
$j^{{\rm th}}$ particle is ${\bf F}_j=-\nabla_j \Psi(\rr^N,t)$,
($\rr^N \equiv \{\rr_1,\rr_2...\rr_N\}$ is the set of position coordinates for
the $N$ particles), so that the system is prevented from relaxing to its
equilibrium distribution. The non-equilibrium probability density function in
this situation will be:
\begin{equation}
P(\rr^N,t) \, =\, \frac{1}{Z} \exp[-\beta \Psi(\rr^N,t)-\beta U(\rr^N,t)]
\label{eq:pdf_1}
\end{equation}
where $Z$ is a normalisation factor, $\beta=1/k_BT$ is the inverse temperature
and $U(\rr^N,t)$ is the potential
energy due to the interparticle interactions and any external potentials. Taking
the gradient of Eq.\ (\ref{eq:pdf_1}) we obtain:
\begin{equation}
{\bf F}_j\,=\, \nabla_j U(\rr^N,t) \,+\, k_BT \frac{\nabla_j
P(\rr^N,t)}{P(\rr^N,t)}.
\label{eq:pdf_2}
\end{equation}
If ${\bf F}_j$ is switched off, there will be a force $-{\bf F}_j$ driving
the diffusion of particle $j$. We now assume that for time scales $\gg \tau_B$,
the Brownian time scale, the velocity of the $i^{{\rm th}}$ particle is
\begin{equation}
{\bf v}_i \,=\, -\sum_{j=1}^N {\bf \Gamma}_{ij} . {\bf F}_j
\label{eq:vel_i}
\end{equation}
where ${\bf \Gamma}_{ij}=\beta{\bf D}_{ij}(\rr^N,t)$ is the mobility tensor and
${\bf D}_{ij}$ is the diffusion tensor. It is implicitly assumed that as
the particles interact, the momentum degrees of freedom equilibrate much faster
than the positional degrees of freedom, and we have effectively averaged over
the momentum degrees of freedom, whilst keeping the particle coordinates fixed.
For a colloidal fluid, this thermal equilibration should occur via the solvent,
and this approximation should be a good one to make. For atomic fluids, this may
not be the case, especially for particles interacting via harshly repulsive
(hard-sphere like) potentials. For fluids interacting with softer potentials,
such as in the Gaussian core model \cite{joe:christos}, this might be
a reasonable approximation to make, particularly when the fluid is at high
densities where each particle interacts with a large number of neighbours
and so the momentum degrees of freedom can equilibrate faster.
Since the particle number is conserved, we can expect the fluid to obey the
continuity equation:
\begin{equation}
\frac{\partial P(\rr^N,t)}{\partial t} \, =\, - \sum_{i=1}^N \nabla_i . [{\bf
v}_i P(\rr^N,t)].
\label{eq:continuity}
\end{equation}
Substituting Eqs.\ (\ref{eq:pdf_2}) and (\ref{eq:vel_i}) into
(\ref{eq:continuity}), one finds
\begin{eqnarray}\notag
\frac{\partial P(\rr^N,t)}{\partial t} \, =\, \sum_{i=1}^N \sum_{j=1}^N \nabla_i
. {\bf \Gamma}_{ij} . \Big[k_BT \nabla_j \hspace{1cm} \\ 
\hspace{1cm} + \, \nabla_j U(\rr^N,t) \Big] P(\rr^N,t).
\label{eq:Smol}
\end{eqnarray}
If the potential energy term $U(\rr^N,t)=0$,
then ${\bf \Gamma}_{ij}=\Gamma \delta_{ij}=\beta D \delta_{ij}$, where $D$ is
the diffusion coefficient and Eq.\ (\ref{eq:Smol}) reduces to
the diffusion equation, $(\partial/ \partial t)P(\rr^N,t) \, =\, \beta \Gamma
\sum_i \nabla_i^2 P(\rr^N,t)$. For a system of interacting particles the
diffusion tensor does not, in general, take such a simple form
\cite{VerbergetalPRE2000}. However, if we neglect the hydrodynamic interactions,
we can replace ${\bf \Gamma}_{ij}$ by its `mean-field' value,
$\Gamma \delta_{ij}$, and then Eq.\ (\ref{eq:Smol}) reduces to a generalised
diffusion equation, termed the Smoluchowski equation:
\begin{equation}
\frac{\partial P(\rr^N,t)}{\partial t} \, =\, \Gamma \sum_{i=1}^N \nabla_i .
[k_BT \nabla_i \, + \, \nabla_i U(\rr^N,t)] P(\rr^N,t).
\label{eq:Smoluch}
\end{equation}
More formally, the Smoluchowski equation is the Fokker-Planck equation for a
system of $N$ Brownian particles in the large friction limit
\cite{Pusey,Risken,Gardiner}. The Langevin equation for a system of $N$ Brownian
particles of mass $m$ is:
\begin{equation}
m \frac{\dd^2 \rr_i(t)}{\dd t^2} \,+\, \Gamma^{-1}\frac{\dd
\rr_i(t)}{\dd t} \, =\, -\nabla_i U(\rr^N,t) \,+\, {\bf X}_i(t),
\label{eq:Langevin1}
\end{equation}
where ${\bf X}_i(t)=(\xi_i^x(t),\xi_i^y(t),\xi_i^z(t))$ is a white noise term
with the property $\left< \xi_i^{\alpha}(t) \right> =0$ and
$\left< \xi_i^{\alpha}(t)\xi_i^{\nu}(t')\right> = 2 k_BT \delta_{ij}
\delta^{\alpha \nu} \delta(t-t')$. When the friction constant $\Gamma^{-1}$ is
large, we may neglect the second derivative with respect to time in Eq.\
(\ref{eq:Langevin1}), and we obtain the stochastic equation of motion
\cite{Marconi:TarazonaJCP1999,Marconi:TarazonaJPCM2000}:
\begin{equation}
\frac{\dd \rr_i(t)}{\dd t}
\, =\, -\Gamma\nabla_i U(\rr^N,t) \,+\, \Gamma{\bf X}_i(t).
\label{eq:Langevin2}
\end{equation}
The (generalised) Fokker-Planck equation for the distribution function
$P(\rr^N,t)$ corresponding to this Langevin equation is Eq.\ (\ref{eq:Smoluch})
\cite{Pusey,Risken,Gardiner}.

\section{Dynamics of the one-body density profile}
\label{sec:2}
In this section we derive an equation for the time evolution of the
one-body density profile, $\rho(\rr,t)$, from the Smoluchowski equation, Eq.\
(\ref{eq:Smoluch}). For a similar approach based solely on pair potentials
see Refs.\
\cite{DhontJCP1996,DhontetalLangmuir1992}. The one body density is merely the
integral of the probability distribution function:
\begin{equation}
\rho(\rr_1,t) \, =\, N \int \dr_2 \, ... \int \dr_N
P(\rr^N,t)
\label{eq:rho_1}
\end{equation}
Similarly, the two-body density is
\begin{equation}
\rho^{(2)}(\rr_1,\rr_2,t) \, =\, N (N-1) \int \dr_3 \, ...\int \dr_N P(\rr^N,t),
\label{eq:rho_2}
\end{equation}
and in general the $n$-particle density is
\begin{equation}
\rho^{(n)}(\rr^n,t) \, =\, \frac{N!}{(N-n)!} \int \dr_{n+1} \,
...\int \dr_N P(\rr^N,t).
\label{eq:rho_n}
\end{equation}
Using Eqs.\ (\ref{eq:rho_1})-(\ref{eq:rho_n}), and assuming that the
potential energy function can be expressed in terms of a one-body external
potential acting on each particle, $V_{ext}(\rr_i,t)$, and that the particle
interactions are a sum of pair potentials, $v_2(\rr_i,\rr_j)$, three body
potentials $v_3(\rr_i,\rr_j,\rr_k)$, and higher body interactions:
\begin{eqnarray} \notag
U(\rr^N,t) \, =\, \sum_{i=1}^N V_{ext}(\rr_i,t) \,
+\, \frac{1}{2} \sum_{j \neq i} \sum_{i=1}^N v_2(\rr_i,\rr_j) \\
+\, \frac{1}{6} \sum_{k \neq j \neq i}\sum_{j \neq i} \sum_{i=1}^N
v_3(\rr_i,\rr_j,\rr_k) \, +\, ...
\label{eq:U_fn}
\end{eqnarray}
then we find that on integrating Eq.\ (\ref{eq:Smoluch}), one obtains
\begin{eqnarray}\notag
\Gamma^{-1} \frac{\partial \rho(\rr_1,t)}{\partial t} \, =\, k_BT \nabla_1^2
\rho(\rr_1,t)\hspace{4cm}\\
+\, \nabla_1 .[\rho(\rr_1,t) \nabla_1 V_{ext}(\rr_1,t)]\hspace{2cm} \notag \\
+\, \nabla_1 . \int \dr_2 \rho^{(2)}(\rr_1,\rr_2,t) \nabla_1
v_2(\rr_1,\rr_2) \notag \\
+\, \nabla_1 . \int \dr_2 \int \dr_3
\rho^{(3)}(\rr_1,\rr_2,\rr_3,t) \nabla_1 v_3(\rr_1,\rr_2,\rr_3) \notag \\
+ \, ....\hspace{4cm}
\label{eq:Dhont_eq}
\end{eqnarray}
We note that if $\partial \rho(\rr,t)/ \partial t=0$, then Eq.\
(\ref{eq:Dhont_eq}) is equivalent to the derivative of the first equation of the
YBG hierarchy \cite{HM}.

For a fluid in {\em equilibrium}, there is an exact sum rule \cite{Evans79}
which relates the gradient of the one body direct correlation function to the
interparticle forces acting on a particle (recall
that $-k_BT c^{(1)}(\rr)$ is the effective one-body potential due to
interactions in the fluid).
If the particles interact solely via pair potentials:
\begin{equation}
- \, k_BT \rho(\rr) \nabla c^{(1)}(\rr) \, =\, \int \dr'
\rho^{(2)}(\rr,\rr') \nabla v_2(\rr,\rr').
\label{eq:grad_c1}
\end{equation}
This result can be generalised straightforwardly to fluids where the particles
interact via many-body potentials, Eq.\ (\ref{eq:U_fn}), giving:
\begin{eqnarray}\notag
- \, k_BT \rho(\rr_1) \nabla c^{(1)}(\rr_1) \,= \hspace{4cm} \\
\sum_{n=2}^{\infty} \int \dr_2 \, ... \int \dr_n \rho^{(n)}(\rr^n)
\nabla_1 v_n(\rr^n).\,\,
\label{eq:grad_c1_many}
\end{eqnarray}
From equilibrium statistical mechanics one also knows that $c^{(1)}(\rr)$ is
equal to the functional derivative of the excess (over ideal) part of the
Helmholtz free energy functional \cite{Evans79}:
\begin{equation}
c^{(1)}(\rr) \, = \, -\beta
\frac{\delta F_{ex}[\rho(\rr)]}{\delta \rho(\rr)},
\label{eq:c1}
\end{equation}
evaluated at the equilibrium density. (Generally $c^{(1)}(\rr)$ is a functional
of $\rho(\rr)$.) Making the {\em approximation} that these identities, Eqs.\
(\ref{eq:grad_c1_many}) and (\ref{eq:c1}), valid for the equilibrium fluid, hold
also for the non-equilibrium fluid and substituting into Eq.\
(\ref{eq:Dhont_eq}), we obtain the DDFT equation, stated without justification
by Evans (see Eqs.\ (166) and (167) of Ref.\ \cite{Evans79}), and derived much
more convincingly by MT
\cite{Marconi:TarazonaJCP1999,Marconi:TarazonaJPCM2000}, i.e.\
\begin{equation}
\Gamma^{-1} \frac{\partial \rho(\rr,t)}{\partial t} \, =\, \nabla .
\left[ \rho(\rr,t) \nabla \frac{\delta F[\rho(\rr,t)]}{\delta
\rho(\rr,t)} \right]
\label{eq:mainres}
\end{equation}
where  $F[\rho(\rr,t)]$ is the Helmholtz free energy functional:
\begin{eqnarray} \notag
F[\rho(\rr,t)] \,=\, k_BT \int \dr \rho(\rr,t)[\ln(\rho(\rr,t)\Lambda^3)-1] \\
+\, F_{ex}[\rho(\rr,t)] \,+ \, \int \dr V_{ext}(\rr,t) \rho(\rr,t).
\label{eq:freeenergy}
\end{eqnarray}
The first term is the ideal gas free energy; $\Lambda$ is the de Broglie
wavelength. In obtaining Eq.\ (\ref{eq:mainres}) by using Eqs.\
(\ref{eq:grad_c1_many}) and (\ref{eq:c1}), which are strictly
equilibrium results, we are effectively assuming that the correlations between
the particles when the fluid is out of equilibrium are equivalent to those for
an equilibrium fluid with the {\it same} one-body density profile $\rho(\rr,t)$.

We can obtain further insight into the status of Eq.\ (\ref{eq:mainres}) by
re-writing it as follows:
\begin{equation}
\Gamma^{-1} \frac{\partial \rho(\rr,t)}{\partial t} \, =\, 
\nabla . [ \, \rho(\rr,t) \nabla \mu(\rr,t) \, ],
\label{eq:Smoluchowski}
\end{equation}
where $-\nabla \mu(\rr,t) \equiv -\nabla (\delta F[\rho]/ \delta \rho)$ is the
net driving force acting on a particle located at $(\rr,t)$. The chemical
potential obtained from Eq.\ (\ref{eq:freeenergy}) has three contributions:
\begin{equation}
\frac{\delta F[\rho(\rr,t)]}{\delta \rho(\rr,t)} \, \equiv \,
\mu(\rr,t) \, =\, \mu_{id}(\rr,t) \,+\,
\mu_{int}(\rr,t) \,+\, \mu_{ext}(\rr,t).
\end{equation}
The first contribution is the ideal gas entropic term $\mu_{id}(\rr,t)= k_BT \ln
\Lambda^3 \rho(\rr,t)$, the second contribution, $\mu_{int}(\rr,t)=-k_BT
c^{(1)}(\rr)$, is that due to the interactions with the other particles in the
fluid, and the final term is simply the external potential,
$\mu_{ext}(\rr,t)= V_{ext}(\rr,t)$. As noted by MT
\cite{Marconi:TarazonaJCP1999}, for non-interacting particles, $\mu_{int} \equiv
0$, and Eq.\ (\ref{eq:Smoluchowski}) reduces to the exact equation for the
diffusion of an ideal Brownian gas, i.e.
\begin{equation}
\Gamma^{-1} \frac{\partial \rho(\rr,t)}{\partial t} \, =\, 
k_BT \nabla^2  \, \rho(\rr,t)
\,+\, \nabla . [\rho(\rr,t) \nabla V_{ext}(\rr,t) \,].
\label{eq:Smoluchowski_ideal_gas}
\end{equation}
For an inhomogeneous interacting fluid in equilibrium the density profile
satisfies \cite{Evans79}
\begin{equation}
\frac{\delta F[\rho(\rr)]}{\delta \rho(\rr)} \, = \,
\mu \, = \, {\rm constant},
\label{eq:ELeq}
\end{equation}
where the Helmholtz free energy
functional $F[\rho]$ is given by Eq.\ (\ref{eq:freeenergy}) with $\rho(\rr,t)$
replaced by $\rho(\rr)$ and $V_{ext}(\rr,t)$ replaced by $V_{ext}(\rr)$, so
there is no net driving force on the particles. It follows that
for a time independent external
potential: $V_{ext}(\rr,t) \rightarrow V_{ext}(\rr)$ as $t \rightarrow \infty$,
regardless of the initial profile $\rho(\rr,t=0)$,
Eq.\ (\ref{eq:mainres}) will yield the same one-body density profile in the
limit $t \rightarrow \infty$ as does
the equilibrium DFT (i.e.\ the solution to Eq.\ (\ref{eq:ELeq})) for the
same external potential. The present derivation
provides, we believe, additional insight to that of MT
\cite{Marconi:TarazonaJCP1999,Marconi:TarazonaJPCM2000} into the physics
incorporated into Eq.\ (\ref{eq:mainres}) (the key DDFT equation). Assuming one
has an accurate expression for
the Helmholtz free energy functional Eq.\ (\ref{eq:freeenergy}), and in
particular for the excess Helmholtz free energy functional,
Eq.\ (\ref{eq:mainres}) should
provide an accurate description of the dynamics of $\rho(\rr,t)$ for a system of
Brownian particles.

\section{Spinodal decomposition from the DDFT equation}
\label{sec:3}

In this section we apply the DDFT derived in the previous section to fluid
spinodal decomposition. Since the basis for the DDFT is the Smoluchowski
equation, an equation of motion for Brownian particles, we expect our theory to
be particularly relevant to spinodal decomposition in colloidal fluids, rather
than molecular fluids. In colloidal fluids friction with the solvent
results in a much faster equilibration of the momentum degrees of freedom
compared with those of the particle positions. However, since we do not
explicitly include the particles of the solvent in which the
colloids are suspended, our theory neglects the hydrodynamic interactions that
may be significant to spinodal decomposition in some colloidal fluids.

When a (colloidal) fluid which exhibits liquid-gas phase separation (or more
generally fluid-fluid phase separation) is quenched to a
state point in the region of the phase diagram where there is coexistence, the
fluid can phase-separate in two distinct ways. The first mechanism is that
which occurs when the state point to which the fluid is quenched is near to the
binodal. In this case phase separation generally occurs via {\em nucleation} of
droplets of one phase forming in the other phase \cite{Gunton,Onuki}. For
example, if the fluid is quenched to a state point inside the binodal on the
liquid side, then bubbles of the gas phase can nucleate in the bulk of the
metastable liquid.

However, if the fluid is quenched to a state point well inside the binodal, then
a different mechanism for phase separation is possible: spinodal decomposition.
Spinodal decomposition is characterised by the exponential growth of density
fluctuations of certain wavelengths \cite{Gunton,Onuki}. In (mean-field)
theoretical descriptions, liquid-gas phase separation is determined by the
occurrence of a van der Waals loop in the Helmholtz free energy per particle,
$f(v)$, where $v$ is the volume per particle. Spinodal decomposition is
predicted to occur in regions of the phase diagram where $(\partial^2 f/\partial
v^2)_T < 0$, i.e.\ regions where the isothermal compressibility $\chi_T$ is
predicted to be negative. The boundary to this region, the spinodal, is defined
by the locus of $(\partial^2 f/\partial v^2)_T = 0$ in the phase diagram.
Experimentally there is not necessarily a sharp distinction between regions
where phase separation occurs via nucleation and via spinodal decomposition.
However, in a deep quench far from the spinodal, spinodal decomposition is the
mechanism generally expected for phase separation.

In a fluid undergoing spinodal decomposition three different regimes can be
distinguished. For early times after the quench, the amplitude of the density
fluctuations are small and theories linear in the density fluctuations such as
the well known Cahn-Hilliard theory \cite{CahnHilliard,Cahn} (see also Refs.\
\cite{Gunton,Onuki,DhontJCP1996,Evans:TDGammaMolecP1979,AbrahamJCP1976})
provide a good
description of this (early) stage of spinodal decomposition. At intermediate
times the density fluctuations can be large, but sharp interfaces between
domains of gas-like and liquid-like regions have still not formed
\cite{DhontJCP1996}. At later stages there are sharp interfaces between domains
of liquid and gas and successful theoretical descriptions of the later stage
dynamics of spinodal decomposition, such as the
Allen-Cahn theory \cite{Allen-Cahn} (see also Refs.\ \cite{Bray,Gunton,Onuki}),
focus on the dynamics of the interfaces. 

First, in Sec.\ \ref{subsec:early_times}, we shall use the DDFT formalism to
derive a generalisation of the (linear)
Cahn-Hilliard theory \cite{CahnHilliard,Cahn}, similar to that described
in Refs.\ \cite{DhontJCP1996,Evans:TDGammaMolecP1979,AbrahamJCP1976},
for the early stages of spinodal decomposition, when the density fluctuations
are small. Then in Sec.\ \ref{subsec:int_times} we will proceed to derive a
non-linear theory which we believe may be applicable
for the dynamics of spinodal decomposition at both short and
intermediate time scales. Results of explicit calculations for a model fluid are
given in Sec.\ \ref{sec:Results_for_a_model_fluid}.

\subsection{Early stages of spinodal decomposition}
\label{subsec:early_times}

We consider spinodal decomposition in the bulk of a fluid, so we set
$V_{ext}(\rr,t)=0$ in Eqs.\ (\ref{eq:mainres}) and (\ref{eq:freeenergy}) and
consider small density fluctuations $\tilde{\rho}(\rr,t)=\rho(\rr,t)-\rho_b$
about the bulk fluid
density, $\rho_b$, i.e.\ we are considering a homogeneous fluid
which has been rapidly quenched to the region of the phase diagram inside
the spinodal (e.g.\ from A to B in Fig.\ \ref{fig:phase_diag}) and we seek those
wavenumbers $k$ for which density fluctuations grow. From
Eqs.\ (\ref{eq:c1}), (\ref{eq:mainres}) and (\ref{eq:freeenergy}) we obtain:
\begin{eqnarray}\notag
(\Gamma k_BT)^{-1} \frac{\partial \tilde{\rho}(\rr,t)}{\partial t}
\, =\, \nabla^2 \tilde{\rho}(\rr,t)
\,-\, \rho_b \nabla^2 c^{(1)}(\rr,t) \\
-\, \nabla . [ \, \tilde{\rho}(\rr,t) \nabla c^{(1)}(\rr,t) \, ].
\label{eq:Smoluchowski_2}
\end{eqnarray}
This approach is basically that of Refs.\
\cite{Evans79,Evans:TDGammaMolecP1979}.
In Ref.\ \cite{Evans79} Evans writes down Eq.\
(\ref{eq:mainres}), and then linearises Eq.\ (\ref{eq:Smoluchowski_2}) in
$\tilde{\rho}$ by Taylor expanding $c^{(1)}$ about the bulk fluid value, giving
\begin{equation}
c^{(1)}(\rr) \, =\, c^{(1)}(\infty)
\,+\, \int \, \dr' \frac{\delta c^{(1)}(\rr)}{\delta
\rho(\rr')} \Bigg \vert_{\rho_b} \tilde{\rho}(\rr',t)
\, +\, {\cal O} (\tilde{\rho}^2),
\label{eq:c_1_expansion}
\end{equation}
where $c^{(1)}(\infty) \equiv c^{(1)}[\rho_b]=-\beta \mu_{ex}$ and
$\mu_{ex}$ is the excess chemical potential.
The second term simplifies by recalling \cite{Evans79}
\begin{eqnarray}\notag
\frac{\delta c^{(1)}(\rr)}{\delta \rho(\rr')}
&=&-\beta \frac{\delta^2 F_{ex}[\rho]}{\delta \rho(\rr') \delta \rho(\rr)}\\
&=& c^{(2)}(\rr,\rr')\notag \\
&=& c^{(2)}(|\rr-\rr'|;\rho_b) ,
\label{eq:c_2}
\end{eqnarray}
for a homogeneous fluid of spherically symmetric particles. For an equilibrium
system $c^{(2)}(r;\rho_b)$ is the Ornstein-Zernike pair direct correlation
function of the fluid of density $\rho_b$. Substituting
Eq.\ (\ref{eq:c_1_expansion}) into Eq.\ (\ref{eq:Smoluchowski_2}), keeping only
terms that are linear in the fluctuation $\tilde{\rho}$, we
obtain \cite{Evans79,Evans:TDGammaMolecP1979}:
\begin{eqnarray}\notag
(\Gamma k_BT)^{-1} \frac{\partial \tilde{\rho}(\rr,t)}{\partial t}
\, =\, \nabla^2 \tilde{\rho}(\rr,t)\hspace{2.5cm} \\
-\, \rho_b \nabla^2 [ \, \int \, \dr' c^{(2)}(|\rr-\rr'|;\rho_b)
\tilde{\rho}(\rr',t) \, ].
\label{eq:Spin_dec_eq}
\end{eqnarray}
Fourier transforming yields an
equation for the time evolution of the different Fourier components
\begin{equation}
\hat{\rho}(\kk,t) \, =\, \int \dr \exp(i \kk.\rr) \tilde{\rho}(\rr,t),
\end{equation}
and one obtains
\begin{equation}
(\Gamma k_BT)^{-1} \frac{\partial \hat{\rho}(k,t)}{\partial t}
\, =\, -k^2 \hat{\rho}(k,t)
\,+\, \rho_b k^2 \, \hat{c}(k) \hat{\rho}(k,t),
\label{eq:Spin_dec_eq_2}
\end{equation}
where $\hat{c}(k)=\int \dr \exp(i \kk.\rr) c^{(2)}(r;\rho_b)$.
The solution of Eq.\ (\ref{eq:Spin_dec_eq_2}) is
\begin{equation}
\hat{\rho}(k,t) \, =\, \hat{\rho}(k,0) \exp [R(k) t],
\label{eq:short_time_law}
\end{equation}
where $R(k)=-\Gamma k_BT \, k^2 (1 \,- \, \rho_b \hat{c}(k))$. For an
{\em equilibrium} fluid, at a state point outside the spinodal, $S(k) \equiv (1
\,- \, \rho_b \hat{c}(k))^{-1}$ is the static
structure factor \cite{HM} and, since for an equilibrium fluid
$S(k)>0$ for all values of $k$, it follows that
outside the spinodal $R(k) <0$ for all values of $k$. From
Eq.\ (\ref{eq:short_time_law}), all Fourier components will
decay implying, of course, the fluid is stable.

On approaching the spinodal from the single phase region
one finds that $S(k=0) \rightarrow \infty$; at the spinodal $(1 \,- \,
\rho_b \hat{c}(k=0))=0$. Within the mean-field (van der Waals-like) theory of
fluids to be described below, we find that inside the spinodal the quantity
$R(k)$ can have positive values for
$k < k_c$, where the value of $k_c$ depends upon how far into the spinodal
region one has quenched, see for example Fig.\ \ref{fig:SofkatB}, which employs
a particular approximation, namely Eq.\ (\ref{eq:hat_c2_fluid}), for
$\hat{c}(k)$.
$k_c$ is obtained as the solution to the equation $\rho_b
\hat{c}(k_{c})=1$. Thus we find that inside the spinodal region there will be
some density fluctuations with a wave number $k<k_{c}$
whose amplitude will grow exponentially
\cite{CahnHilliard, Cahn, Evans79, Evans:TDGammaMolecP1979, DhontJCP1996,
DhontetalLangmuir1992, AbrahamJCP1976}. The deeper the quench into the spinodal
region, the larger the value of $k_c$ can be.

This general picture of the exponential growth of certain Fourier components
(modes) was obtained originally by Cahn and Hilliard
\cite{CahnHilliard,Cahn} who derived an
explicit approximation for the function $R(k)$. Cahn-Hilliard theory for
spinodal decomposition is usually derived
by considering the continuity equation (\ref{eq:cont_eq}), together with the
following approximation for the current
\begin{equation}
\mathbf{j}(\rr,t)=-\, M \nabla \frac{\delta F[\rho(\rr,t)]}{\delta \rho(\rr,t)},
\label{eq:current_CH}
\end{equation}
where $M$ is a mobility constant and the functional $F$ is chosen to be of the
$\phi^4$ Ginzburg-Landau form. Generalising slightly we assume that
the free energy functional has the square-gradient form \cite{Evans92}:
\begin{equation}
F_{sg}[\rho(\rr,t)]\, =\,
\int\dr\, \bigg[ f_0(\rho(\rr,t)) \,+\,
\frac{1}{2}K|\nabla \rho(\rr,t)|^2 \bigg],
\label{eq:F_GL}
\end{equation}
where $f_0(\rho) \equiv \rho f(\rho)$ is the Helmholtz free energy density for
the homogeneous fluid of density $\rho$ and we treat $K$ as a positive
constant. From Eq.\ (\ref{eq:cont_eq}) one then finds
\begin{equation}
\frac{\partial \rho(\rr,t)}{\partial t} \, =\, M \nabla^2 \bigg[ \frac{\partial
f_0(\rho(\rr,t))}{\partial \rho(\rr,t)} \,-\, K \nabla^2 \rho(\rr,t) \bigg].
\label{eq:Cahn-Hilliard_1}
\end{equation}
This equation is then linearised about the bulk density $\rho_b$ to obtain
\cite{Gunton,Evans:TDGammaMolecP1979}:
\begin{equation}
\frac{\partial \tilde{\rho}(\rr,t)}{\partial t} \, =\, M \nabla^2 \bigg[
\left( \frac{\partial^2 f_0}{\partial \rho^2}\right)_{\rho_b}
-\, K \nabla^2 \bigg] \tilde{\rho}(\rr,t).
\label{eq:Cahn-Hilliard_2}
\end{equation}
On Fourier transforming Eq.\ (\ref{eq:Cahn-Hilliard_2}) one obtains $\partial
\hat{\rho}(k,t)/\partial t = R_{sg}(k) \hat{\rho}(k,t)$, where
$R_{sg}(k)=-Mk^2(Kk^2+(\partial^2 f_0/\partial \rho^2)_{\rho_b})$,
the solution to
which is Eq.\ (\ref{eq:short_time_law}), with $R(k)$ replaced by $R_{sg}(k)$.
Note that $(\partial^2 f_0/\partial \rho^2)_{\rho_b}$ is negative inside the
spinodal. It is clear from this argument that
Cahn-Hilliard theory can therefore be viewed as a special case of the more
general linear theory presented in the earlier part of this subsection
\cite{footnote2}. By employing a free energy functional more accurate than
(\ref{eq:F_GL}) one should be able to incorporate short wavelength density
fluctuations as well as the long wavelength, $k \rightarrow 0$, fluctuations
accounted for by a square-gradient approach.

\subsection{Spinodal decomposition at intermediate times}
\label{subsec:int_times}

In order to go one step beyond the lowest order (linear) theories described
above, we consider an
approximate excess Helmholtz free energy functional, obtained by Taylor
expanding the excess Helmholtz free energy about the uniform density. By
integrating Eq.\ (\ref{eq:c_1_expansion}) and omitting terms beyond
${\cal O}(\tilde{\rho}^2)$ we obtain
\begin{eqnarray}\notag
F_{ex}[\rho] \,=\, F_{ex}[\rho_b] \,+\, \mu_{ex}
\int \dr \tilde{\rho}(\rr,t) \hspace{2cm} \\
-\, \frac{k_BT}{2} \int \dr \int \dr'
\tilde{\rho}(\rr,t)\tilde{\rho}(\rr',t) c^{(2)}(|\rr-\rr'|; \rho_b).
\label{eq:F_ex}
\end{eqnarray}
This truncated quadratic density expansion is often used in the theory of
inhomogeneous fluids in equilibrium \cite{Evans92}. Using Eq.\
(\ref{eq:F_ex}) in Eq.\ (\ref{eq:Smoluchowski_2}) we find:
\begin{eqnarray}\notag
(\Gamma k_BT)^{-1} \frac{\partial \tilde{\rho}(\rr,t)}{\partial t} \, =\,
\nabla^2 \tilde{\rho}(\rr,t) \\
-\, \nabla . \bigg[ (\rho_b \,+\,
\tilde{\rho}(\rr,t)) \int \dr' \tilde{\rho}(\rr',t) \nabla
c^{(2)}(|\rr-\rr'|;\rho_b)
\bigg]. \,\,\,\,
\label{eq:abc}
\end{eqnarray}
The first two terms on the right hand side are those entering the linear theory
(\ref{eq:Spin_dec_eq}) while the third term is the only non-linear one which
arises for the functional (\ref{eq:F_ex}).
Fourier transforming Eq.\ (\ref{eq:abc}) we obtain:
\begin{eqnarray}\notag
(\Gamma k_BT)^{-1}  \frac{\partial \hat{\rho}(k,t)}{\partial t}
\, =\, -k^2 \hat{\rho}(k,t)
\,+\, \rho_b k^2 \, \hat{c}(k) \hat{\rho}(k,t) \\
+\, \frac{1}{(2 \pi)^3} \int \dk' \, \kk.\kk'
\hat{\rho}(k',t)\hat{c}(k')\hat{\rho}(|\kk-\kk'|,t), \,\,
\label{eq:Spin_dec_eq_2_next}
\end{eqnarray}
which should be compared with Eq.\ (\ref{eq:Spin_dec_eq_2}); there an additional
term on the right hand side that is non-linear in $\hat{\rho}(k,t)$. This term
describes the coupling between the different Fourier components of the density
fluctuations (modes) \cite{footnote1}.
An equation almost equivalent to Eq.\ (\ref{eq:Spin_dec_eq_2_next})
was derived recently by considering the mobility $M$ to be a linear
function of the order parameter, rather than a constant, in a
Cahn-Hilliard treatment -- see Eq.\ (10) in Ref.\ \cite{MaoetalEurPhysJE2001};
we shall discuss this further in Sec.\ \ref{sec:spin_discuss}. In order
to proceed further we must assume a particular form for the Helmholtz free
energy functional, from which we can obtain $\hat{c}(k)$ and thus solve
Eq.\ (\ref{eq:Spin_dec_eq_2_next}) numerically.

\subsection{Results for a model fluid}
\label{sec:Results_for_a_model_fluid}

\begin{figure}
\includegraphics[width=8cm]{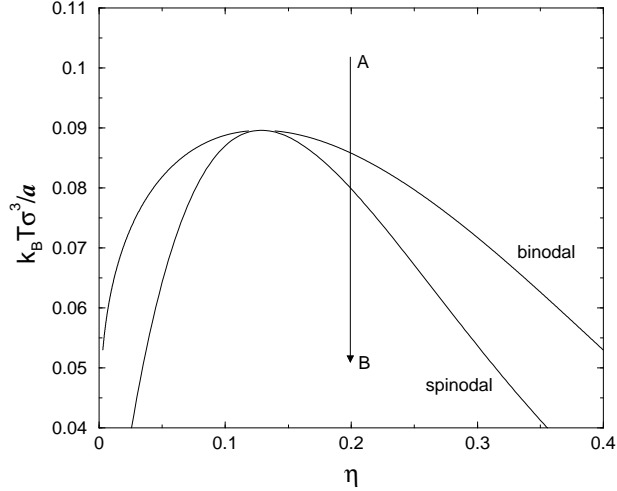}
\caption{\label{fig:phase_diag}
Phase diagram for a fluid composed of particles interacting via pair
potentials of the form in Eqs.\ (\ref{eq:pair_pot}-\ref{eq:pair_pot_at}),
calculated from the free-energy of Eq.\ (\ref{eq:freefunc_HS_ref}).
$\eta=\pi \rho_b \sigma^3/6$ is the
packing fraction and $k_BT \sigma^3/a$ is the (reduced) temperature.
The path from $A$ to $B$ denotes the quench corresponding to
the results in Figs.\ \ref{fig:SofkatB} and \ref{fig:hatrho}.}
\end{figure}

We consider a fluid composed of particles interacting via pair potentials of the
form
\begin{equation}
v_2(r)\,=\, v_{hs}(r)\,+\,v_{at}(r),
\label{eq:pair_pot}
\end{equation}
where
\begin{equation}
v_{hs}(r)\, = 
\begin{cases}
\infty \hspace{8mm} r \leq \sigma \\
0 \hspace{10mm} r > \sigma
\end{cases} 
\label{eq:pair_pot_hs}
\end{equation}
is the hard-sphere pair potential, and the attractive part of the pair
potential has a Yukawa form:
\begin{equation}
v_{at}(r)\, = \,- \frac{a \lambda^3 \exp(-\lambda r)}{4 \pi \lambda r};
\label{eq:pair_pot_at}
\end{equation}
where $a$ and $\lambda$ are positive constants. Provided the decay length
$\lambda^{-1}$ is sufficiently large this model fluid exhibits stable, with
respect to freezing, liquid-gas phase separation. Pair potentials of this
form are often used as crude models for simple fluids but they could be used to
model the effective (depletion) potential between the colloids in a
colloid-polymer mixture solution \cite{Likos} by choosing the pair potential
parameters in Eqs.\ (\ref{eq:pair_pot_hs}) and (\ref{eq:pair_pot_at}) to mimic
the well-known Asakura-Oosawa potential
\cite{Likos,BraderetalMolecPhys2003,AO1,AO2}. In calculations for our model
system we approximate the excess Helmholtz free energy functional by
\begin{equation}
F_{ex}[\rho] \,=\, F_{ex}^{hs}[\rho] \,+\, \frac{1}{2} \int \dr \int \dr'
\rho(\rr,t)\rho(\rr',t) v_{at}(|\rr-\rr'|),
\label{eq:freefunc_HS_ref}
\end{equation}
where $F_{ex}^{hs}[\rho]$ is the reference hard-sphere Helmholtz excess free
energy functional and attractive interactions are treated in a mean-field
fashion \cite{Evans92}. If we employ the Rosenfeld Fundamental Measure Theory
\cite{RosenfeldPRL1989, Rosenfeld:Levesque:WeisJCP1990, RosenfeldJCP1990}
for $F_{ex}^{hs}[\rho]$, this non-local functional generates the Percus-Yevick
pair direct correlations functions in a hard sphere fluid.
Thus, using Eq.\ (\ref{eq:c_2}) we obtain the following simple (Random Phase)
approximation
\begin{equation}
c^{(2)}(r;\rho_b)\,=\, c_{PY}(r;\rho_b) \, - \, \beta v_{at}(r),
\label{eq:c2_fluid}
\end{equation}
where $c_{PY}(r)$ is the Percus-Yevick (PY) approximation \cite{HM}
for the hard-sphere
pair direct correlation function. With this choice the Fourier
transform of $c^{(2)}(r)$ can be carried out analytically:
\begin{equation}
\hat{c}(k)\,=\, \hat{c}_{PY}(k) \, + \,
\frac{\beta a \lambda^2}{\lambda^2 + k^2},
\label{eq:hat_c2_fluid}
\end{equation}
where $\hat{c}_{PY}(k)$ is given by \cite{Ashcroft:LecknerPhysRev1966}:
\begin{eqnarray}
\hat{c}_{PY}(k)\,=&-&4 \pi \sigma^3 \bigg[ \left(\frac{\alpha+2 \beta+4
\gamma}{q^3}\,-\, \frac{24 \gamma}{q^5}\right) \sin(q) \notag \\
&+& \left( -\frac{\alpha+\beta+\gamma}{q^2} +
\frac{2 \beta + 12 \gamma}{q^4} - \frac{24 \gamma}{q^6} \right)
\cos(q) \notag \\
&+& \left( \frac{24 \gamma}{q^6} - \frac{2 \beta}{q^4} \right) \bigg]
\label{eq:hat_c2_fluid_PY}
\end{eqnarray}
where $q=k \sigma$, and the coefficients
\begin{eqnarray}
\alpha&=&\frac{(1+2\eta)^2}{(1-\eta)^4}, \notag \\
\beta&=&\frac{-6 \eta(1+\eta/2)^2}{(1-\eta)^4}, \notag \\
\gamma&=&\frac{\eta(1+2\eta)^2}{2(1-\eta)^4}
\end{eqnarray}
depend upon $\eta=\pi \rho_b \sigma^3/6$, the packing fraction.
The phase diagram for our model fluid is displayed in Fig.\
\ref{fig:phase_diag}. The liquid-gas binodal is
calculated by performing the common tangent construction on the Helmholtz free
energy per particle for the bulk fluid, $f(\rho_b)=f_{PY}(\rho_b)-\rho_ba/2$,
obtained from Eq.\ (\ref{eq:freefunc_HS_ref});
$f_{PY}(\rho_b)$ is the PY compressibility approximation for the hard-sphere
Helmholtz free energy \cite{HM}. The spinodal is
the locus of points for which $\partial^2 f/ \partial v^2=0$, where
$v=1/\rho_b$. For simplicity we choose to set the inverse length scale parameter
in the attractive part of the pair potential equal to the hard-sphere diameter,
i.e.\ $\lambda^{-1}=\sigma$. The quantity $a=\int \dr \,
v_{at}(r)$ determines the energy scale. We expect the present model fluid to
exhibit a freezing transition for large $\eta$, but we do not consider this
here.

\begin{figure}
\includegraphics[width=8cm]{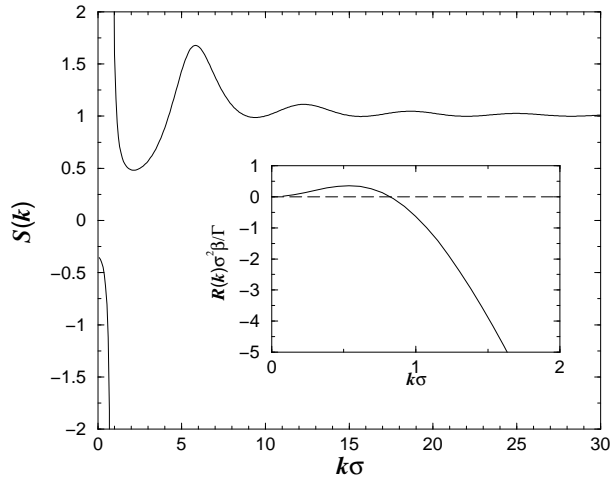}
\caption{\label{fig:SofkatB}
The function $S(k) \equiv (1 \,- \, \rho_b \hat{c}(k))^{-1}$ calculated using
Eq.\ (\ref{eq:hat_c2_fluid})
at the state point with $\eta=0.2$ and $k_BT \sigma^3/a=0.05$, corresponding to
point $B$ in Fig.\ \ref{fig:phase_diag}. Note that for $k<k_c$, where $k_c
\sigma \simeq 0.8$, $S(k) <0$.
In the inset we plot $R(k) \sigma^2 \beta/\Gamma=-k^2 \sigma^2/S(k)$, the factor
appearing in the exponential in Eq.\ (\ref{eq:short_time_law}). In the initial
stages of spinodal decomposition density fluctuations with wave numbers $k<k_c$
grow exponentially, whereas for $k>k_c$ the fluctuations are damped.}
\end{figure}

We are now in a position to use our approximate
form for $\hat{c}(k)$, obtained from Eqs.\ (\ref{eq:hat_c2_fluid}) and
(\ref{eq:hat_c2_fluid_PY}), together with Eq.\ (\ref{eq:Spin_dec_eq_2_next}), to
calculate $\hat{\rho}(k,t)$ during the early and intermediate times of spinodal
decomposition. We {\em assume} that at $t=0$ $\hat{\rho}(k,t)$ takes a small,
constant, positive value for all values of $k$ and choose
$\hat{\rho}(k,t=0)=10^{-8}$. The later time dynamics is insensitive to the
specific choice of value for $\hat{\rho}(k,t=0)$ because for short times, when
Eq.\ (\ref{eq:short_time_law}) describes the dynamics of spinodal decomposition,
Fourier components (modes) with wave number $k>k_c$ do not grow and are
exponentially damped, whereas components with $k<k_c$ grow exponentially with
the initial value
$\hat{\rho}(k,t=0)$ as a prefactor. Thus choosing a different value for
$\hat{\rho}(k,t=0)$ is equivalent to a shift of the time axis. For a quench to
the state point $k_BT \sigma^3/a=0.05$ and $\eta =0.2$ (i.e.\ the quench to
point $B$, in Fig.\ \ref{fig:phase_diag}) the early time
growth gives a single peak in $\hat{\rho}(k,t)$ at $k \sigma \simeq 0.6$
corresponding to where the
maximum of $R(k)$ in Eq.\ (\ref{eq:short_time_law}) occurs -- see the inset to
Fig.\ \ref{fig:SofkatB}. We find that $R(k)$ obtained using Eqs.\
(\ref{eq:hat_c2_fluid}) and (\ref{eq:hat_c2_fluid_PY}) for $\hat{c}(k)$
has a similar form to that extracted from molecular dynamics simulation results
for a Lennard-Jones fluid deep inside the spinodal
region \cite{AbrahametalJCP1979}. Moreover, the overall shape of $R(k)$ and the
values of $k_c \sigma$ are in keeping with results for the Lennard-Jones fluid
(for similar state points) obtained by Evans and Telo Da Gama
\cite{Evans:TDGammaMolecP1979}, using a theory equivalent to the present, and by
Abraham \cite{AbrahamJCP1976}, using a perturbation theory approach.
We also plot in Fig.\ \ref{fig:SofkatB} the function $S(k) \equiv
(1 \,- \, \rho_b \hat{c}(k))^{-1}$; the negative portion of this
function corresponds to wave numbers $k$ for which density fluctuations grow
exponentially in the early stages of spinodal decomposition.

\begin{figure}
\includegraphics[width=8cm]{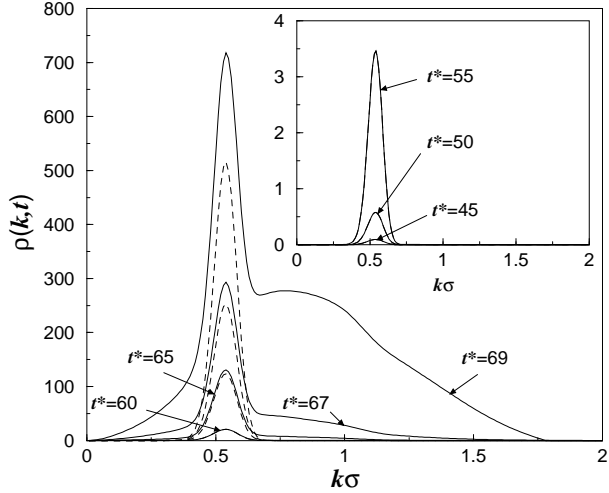}
\caption{\label{fig:hatrho}
Plot of $\hat{\rho}(k,t)$ for a quench to the state point
$k_BT \sigma^3/a=0.05$ and $\eta =0.2$, point $B$ in
Fig.\ \ref{fig:phase_diag}. $\hat{\rho}(k,t)$  is shown for times $t^*=k_BT
\Gamma \sigma^2 t = 45,50,55$ (in the inset) and $t^*=60,65,67,69$ (in the main
figure). The results obtained from the linear theory (Eq.\
(\ref{eq:Spin_dec_eq_2})) are denoted by a dashed line, and those from
the non-linear theory, (Eq.\ (\ref{eq:Spin_dec_eq_2_next})) by a solid line.
The effect of the coupling between Fourier components (modes) with different
wave numbers, described by the non-linear theory,
becomes increasingly significant at later times, whereas for earlier times (see
inset), the results from the two theories are indistinguishable.}
\end{figure}

Some typical plots of $\hat{\rho}(k,t)$ at early times are
displayed in the inset to Fig.\ \ref{fig:hatrho}, for a quench to point $B$
in Fig.\ \ref{fig:phase_diag}. The plots are for the reduced times $t^*=k_BT
\Gamma \sigma^2 t = 45$, 50 and 55 in the inset and 60, 65, 67 and 69 in the
main figure. The results obtained from the linear theory, Eq.\
(\ref{eq:Spin_dec_eq_2}), and the non-linear theory, Eq.\
(\ref{eq:Spin_dec_eq_2_next}), are indistinguishable for the three earliest
times (see inset).
However, at later times the linear theory (dashed line) continues to give just a
single peak in $\hat{\rho}(k,t)$ that grows exponentially, whereas the
non-linear theory(solid line), which includes the effect of coupling between
Fourier components with different wave numbers, shows that components with wave
numbers different from that predicted by the linear theory can also grow. We
see that the effect of the coupling incorporated into the non-linear
theory becomes increasingly significant at intermediate times, producing first a
shoulder which grows into a bump
in $\hat{\rho}(k,t)$ at a larger wave number, $k \sigma \simeq0.8$ than that of
the main peak which first appears in $\hat{\rho}(k,t)$ at early times.

\subsection{Discussion}
\label{sec:spin_discuss}

In simulation studies of spinodal decomposition the quantity that is often
measured in order to characterise the fluid is the structure factor
\begin{equation}
S(k,t) \, =\, \frac{1}{N} \sum_{i,j=1}^N \left< \, \exp(i
\kk.[\rr_i(t)-\rr_j(t)]) \, \right>,
\label{eq:inhom_Sofk}
\end{equation}
where $\left< ... \right>$ is an ensemble average over different realisations of
the stochastic noise in the interval up to time $t$.
For small values of $k$ one finds that \cite{DhontJCP1996}:
\begin{equation}
S(k,t) \, \simeq \, A(t) \, +\, \frac{1}{N} [\hat{\rho}(k,t)]^2,
\label{eq:inhom_Sofk_approx}
\end{equation}
where $A(t)$ is a (small) wave number independent baseline. Thus, to a
reasonable approximation $S(k,t)
\propto [\hat{\rho}(k,t)]^2$ \cite{DhontJCP1996} (see also Ref.\
\cite{Evans:TDGammaMolecP1979}). Simulation studies of the early stages of
spinodal decomposition in model colloidal fluids such as that in Ref.\
\cite{Lodge:HayesJCP1998}, where they consider a fluid composed of particles
interacting via the Lennard-Jones pair potential, display the growth of a single
peak in $S(k,t)$. This is, of course, a general feature of the very early stages
of spinodal decomposition and is found in many other systems
\cite{Gunton,DhontJCP1996}.
Thus, our results for short times showing the growth of a single peak in
$\hat{\rho}(k,t)$, and therefore also in $S(k,t)$, are
in keeping with simulation studies.

The development of a second peak (shoulder)
in $S(k,t)$ at intermediate times after the quench, at a larger value of $k$ was
observed in the two-dimensional
calculations of Mao \etal\ \cite{MaoetalEurPhysJE2001},
where they considered a non-linear extension of Cahn-Hilliard theory for polymer
mixtures. The order parameter for their theory is the deviation of the
concentration $c(\rr,t)=C(\rr,t)-C_0$
(i.e.\ $c(\rr,t)$ replaces $\tilde{\rho}(\rr,t)$ in our
theory). They used the approximation that the mobility
$M=M_0+M_1 c(\rr,t)$, i.e.\ a linear function of $c(\rr,t)$, together with a
square-gradient Helmholtz free energy functional in Eq.\
(\ref{eq:current_CH}). By linearising $\delta F/\delta C$ about the average
concentration $C_0$ in the manner leading to Eq.\ (\ref{eq:Cahn-Hilliard_2})
along with the continuity equation (\ref{eq:cont_eq}), Mao \etal\
\cite{MaoetalEurPhysJE2001} obtain
a non-linear theory that is very similar in structure to
that which we obtain using Eq.\ (\ref{eq:current}) with a
constant mobility together with the truncated functional (\ref{eq:F_ex}).
The main difference between the two approaches, other than the choice
of approximation for the Helmholtz free energy functional \cite{footnote3},
is that Mao \etal\ \cite{MaoetalEurPhysJE2001} have an adjustable parameter, the
ratio $M_1/M_0$, that allows a tuning of the mobility for their polymer blend,
whereas in our approach
we are effectively restricted to the choice $M_1/M_0=1/\rho_b$.
Their results also show that the second peak (shoulder) in $S(k,t)$
results from considering a theory that is non-linear (second order) in the order
parameter. The other significant difference from Ref.\
\cite{MaoetalEurPhysJE2001} is that the present theory uses a
microscopic non-local functional (\ref{eq:F_ex}), and therefore includes the
effects of interparticle correlations more accurately than a gradient
(Ginzburg-Landau) theory
such as that used by Mao \etal\ \cite{MaoetalEurPhysJE2001}. Nevertheless, in
the early and intermediate times of spinodal decomposition, where sharp
interfaces between gas-like and liquid-like domains have
not yet formed, we should expect good qualitative agreement
between our results and those from the non-linear Cahn-Hilliard
gradient theory of Mao \etal\ \cite{MaoetalEurPhysJE2001}.
Mao \etal\ also found that the presence of the
non-linear terms in their theory resulted in a
significant change in the connectivity of contour plots of the order parameter
from that found in the absence of these terms ($M_1=0$) -- see Figs.\ 3, 5 of
Ref.\ \cite{MaoetalEurPhysJE2001}. On the basis of our present analysis
we would expect this observation to be a general feature of spinodal
decomposition at intermediate time scales.

\section{Concluding remarks}
\label{sec:conc}

This paper falls broadly into two parts. In the first part we
provided an alternative derivation of the DDFT developed by MT
\cite{Marconi:TarazonaJCP1999,Marconi:TarazonaJPCM2000}. Our derivation
elucidates the physical approximations made in order to construct the theory.
The starting point is the Smoluchowski equation (\ref{eq:Smoluch}), the
(generalised)
Fokker-Planck equation for the probability distribution function $P(\rr^N,t)$
corresponding to the Langevin equation of motion, Eq.\ (\ref{eq:Langevin2}).
This stochastic basis for the theory means that the theory should be applicable
to colloidal fluids where because of interactions with the solvent particles
the momentum degrees of freedom of the colloids equilibrate much faster than
the positional degrees of freedom. In atomic fluids the equilibration
timescale for the momentum degrees of freedom can be of the same order as that
for the relaxation of positional degrees of freedom and therefore the
present theory may break down for atomic fluids.
However, because the correct equilibrium limit is built into this theory, i.e.\
when $\partial \rho(\rr,t)/ \partial t =0$ the theory is equivalent to Eq.\
(\ref{eq:ELeq}), it is feasible that the present theory
would give reasonable results for atomic fluids, assuming that one also has a
reliable approximation for the excess Helmholtz free energy functional,
$F_{ex}[\rho]$. The fact that the correct equilibrium limit is built in
is, we believe, one of the most appealing features of the theory.

The further approximation (beyond assuming that
the Smoluchowski equation holds)
made in deriving the DDFT, Eq.\ (\ref{eq:mainres}), is to assume Eqs.\
(\ref{eq:grad_c1_many}) and (\ref{eq:c1}), exact equilibrium results, are also
valid for the
non-equilibrium fluid. This approximation is equivalent to assuming that the
two-particle, three-particle and higher order correlations in the
non-equilibrium fluid are equivalent to those in an equilibrium fluid with the
specified one-body density profile. We expect that this approximation is a
reasonable one
to make and will not result in significant failures of the theory, as long as
other approximations can be justified. We therefore believe
that Eq.\ (\ref{eq:mainres}) may well provide a reliable account of the dynamics
for a variety of different fluids, provided of course, that one has an
accurate approximation for the appropriate $F_{ex}[\rho]$.

The second part of the present paper is concerned with
the application of the DDFT
to the problem of spinodal decomposition. Our key result is Eq.\
(\ref{eq:Spin_dec_eq_2_next}). For early times after quenching the fluid, when
density fluctuations are small in amplitude, our theory reduces to the linear
theory of Evans and Telo Da Gama \cite{Evans:TDGammaMolecP1979}. Furthermore,
this linear theory reduces to the Cahn-Hilliard theory if we were to use a
square gradient Helmholtz free energy functional Eq.\ (\ref{eq:F_GL}) rather
than the more accurate non-local functional, Eq.\ (\ref{eq:F_ex}). Cahn-Hilliard
theory \cite{CahnHilliard,Cahn,Gunton,Onuki} is known to
provide a successful qualitative description of the early
stages of spinodal decomposition. Since the present theory
incorporates all the effects that Cahn-Hilliard theory describes, and provides a
more accurate treatment of short wavelength correlations
\cite{Evans:TDGammaMolecP1979} it should be a reliable quantitative theory for
the early stages of spinodal decomposition.

However, the key feature of our theory is that it incorporates a
``mode-coupling'' term (final term on the right hand side of Eq.\
(\ref{eq:Spin_dec_eq_2_next})) which describes the coupling between different
density fluctuation modes, an effect which becomes important in the intermediate
stages of spinodal decomposition. The results of calculations including our
``mode-coupling'' term are in qualitative agreement with
those from a recent non-linear extension to Cahn-Hilliard theory
\cite{MaoetalEurPhysJE2001}. We believe that further
work needs to be done in testing
the predictions of our theory. In particular, a comparison with Brownian
dynamics simulation results for spinodal decomposition would be very useful;
we do not know of relevant simulations which go beyond the very early stages of
spinodal decomposition. Of course, running simulations that get into
the stage of spinodal decomposition for which our results exhibit deviations
from those of the linear theory could be computationally expensive.

We conclude by mentioning one important conceptual issue regarding the input to
our theory of spinodal decomposition. Clearly Eqs.\ (\ref{eq:Spin_dec_eq_2}) or
(\ref{eq:Spin_dec_eq_2_next}) require $\hat{c}(k)$ as input. As emphasised
earlier, for an {\em equilibrium} state $\hat{c}(k)$ is simply the Fourier
transform of the pair direct correlation function and, as such, it can be
obtained from integral equation theories \cite{HM} or from simulations. In
Ref.\ \cite{Evans:TDGammaMolecP1979} results for $\hat{c}(k)$ calculated using
Percus-Yevick theory for a Lennard-Jones fluid were {\em extrapolated} into the
unstable spinodal region \cite{EbneretalPRA1976,TDGamma:EvansMolecP1979}. Such a
procedure is fraught with uncertainty and is difficult to justify. The present
DFT formulation of the theory avoids such problems. $c^{(2)}$ is {\em defined}
as the second functional derivative of $-\beta F_{ex}[\rho]$ and provided one
makes a division of $F_{ex}[\rho]$ into a repulsive reference contribution plus
an attractive (mean-field-like) contribution, as in Eq.\
(\ref{eq:freefunc_HS_ref}), there is no difficulty in calculating
$c^{(2)}(r;\rho_b)$ inside the spinodal. We do not need to make any
extrapolation. One could envisage treating attractive interactions in a more
sophisticated fashion \cite{Evans92} than in Eq.\ (\ref{eq:freefunc_HS_ref}) but
provided the basic division is maintained one should expect to obtain similar
results for $\hat{c}(k)$.

\section*{Acknowledgements}
We are grateful to Christos Likos and Markus Rauscher for useful
discussions and comments on the present work.
A.J.A. acknowledges the support of EPSRC under grant number GR/S28631/01.

\begin{thebibliography}{99}

\bibitem{Evans92}
For example, R. Evans, in {\it Fundamentals of Inhomogeneous Fluids}, ed. D.
Henderson, Dekker, New York, (1992), ch. 3.

\bibitem{Dietrich}
S. Dietrich in {\it Phase Transitions and Critical Phenomena}
{\bf 12}, ed C. Domb and J.L. Lebowitz (London: Academic, 1988) p1.

\bibitem{Marconi:TarazonaJCP1999}
U. Marini Bettolo Marconi and P. Tarazona, {\it J. Chem. Phys.} {\bf 110}
8032 (1999).

\bibitem{Marconi:TarazonaJPCM2000}
U. Marini Bettolo Marconi and P. Tarazona, {\it J. Phys.: Condens. Matter}
{\bf 12} A413 (2000).

\bibitem{Evans79}
R. Evans, {\it Adv. Phys.} {\bf 28} 143 (1979).

\bibitem{Dieterich}
W. Dieterich, H.L. Frisch and A. Majhofer, {\it Z. Phys. B}, {\bf 78}, 317
(1990).

\bibitem{BagchiPhysicaA1987}
B. Bagchi, {\it Physica A} {\bf 145} 273 (1987).

\bibitem{Yoshimori:Day:Patay:JCP1998}
A. Yoshimori, T.J.F. Day and G.N. Patey, {\it J. Chem. Phys.} {\bf 108} 6378
(1998).

\bibitem{joe:christos}
J. Dzubiella and C. N. Likos, {\it J. Phys.: Condens. Matter} {\bf 15}, L147
(2003).

\bibitem{Flor1}
F. Penna and P. Tarazona, {\it J. Chem. Phys.}, {\bf 119}, 1766 (2003).

\bibitem{Flor2}
F. Penna, J. Dzubiella and P. Tarazona, {\it Phys. Rev. E} {\bf 68}, 061407
(2003).

\bibitem{joepreprint}
J. Chakrabarti, J. Dzubiella and H. L\"owen, {\it Pre-print: cond-mat/0403475}

\bibitem{KawasakiPhysicaA1994}
K. Kawasaki, {\it Physica A} {\bf 208} 35 (1994).

\bibitem{Kirkpatrick:ThirumalaiJPhysA1989}
T.R. Kirkpatrick and D. Thirumalai, {\it J. Phys. A} {\bf 22} L149 (1989).

%
\bibitem{Munakata1}
T. Munakata, {\it J. Phys. Soc. Jpn.} {\bf 58} 2434 (1989).

\bibitem{Munakata2}
T. Munakata, {\it J. Phys. Soc. Jpn.} {\bf 59} 1299 (1989).

\bibitem{Munakata3}
T. Munakata, {\it Phys. Rev. E} {\bf 50} 2347 (1994).

\bibitem{DeanJPhysA1996}
D.S. Dean, {\it J. Phys. A} {\bf 29} L613 (1996).

\bibitem{Frusawa:HayakawaJPhysA2000}
H. Frusawa and R. Hayakawa, {\it J. Phys. A}, {\bf 33}, L155 (2000).

\bibitem{KawasakiJPCM2000}
K. Kawasaki,  {\it J. Phys.: Condens. Matter}
{\bf 12} 6343 (2000).

\bibitem{Rauscher}
A.J. Archer and M. Rauscher, {\it to be submitted}.

\bibitem{MunakataPRE2003}
T. Munakata, {\it Phys Rev. E}, {\bf 67}, 022101, (2003).

\bibitem{Pusey}
P.N. Pusey, in {\it Liquids, Freezing and Glass Transition} ed. J-P. Hansen, D.
Levesque and J. Zinn-Justin, North Holland,
Amsterdam (1991).

\bibitem{PuseyTough}
P.N. Pusey and R.J.A. Tough, in {\it Dynamic Light Scattering}, ed. R. Pecora,
Plenum Press, New York, (1985).

\bibitem{Croxton}
C.A. Croxton, {\it Liquid State Physics -- A Statistical Mechanical
Introduction}, Cambridge University Press, Cambridge (1974).

\bibitem{Risken}
H. Risken, {\it The Fokker-Planck Equation: Methods of Solutions and
Applications}, Springer-Verlag, Berlin, (1996), 2nd ed.

\bibitem{Gardiner}
C.W. Gardiner, {\it Handbook of Stochastic Methods for Physics, Chemistry and
the Natural Sciences}, Springer-Verlag, Berlin, (1996).

\bibitem{VerbergetalPRE2000}
R. Verberg, I.M. de Schepper and E.G.D. Cohen, {\it Phys. Rev. E} {\bf 61}
2967 (2000).

\bibitem{DhontJCP1996}
J.K.G. Dhont, {\it J. Chem. Phys.} {\bf 105} 5112 (1996).

\bibitem{DhontetalLangmuir1992}
J.K.G. Dhont, A.F.H. Duyndam and B.J. Ackerson, {\it Langmuir} {\bf 8} 2907
(1992).

\bibitem{HM}
J-P. Hansen and I.R. McDonald, {\it Theory of Simple Liquids}, Academic,
London, (1986), 2nd ed.

\bibitem{Gunton}
J.D. Gunton, M. San Miguel and P.S. Sahni in {\it Phase Transitions and
Critical Phenomena} {\bf 8}, ed C. Domb and J.L. Lebowitz (London: Academic,
1988) p267.

\bibitem{Onuki}
A. Onuki, {\it Phase Transition Dynamics}, Cambridge University Press,
Cambridge, (2002).

\bibitem{CahnHilliard}
J.W. Cahn and J.E. Hilliard, {\it J. Chem. Phys.} {\bf 31}, 688 (1959).

\bibitem{Cahn}
J.W. Cahn, {\it Acta. metall.}, {\bf 9}, 795 (1961).

\bibitem{Evans:TDGammaMolecP1979}
R. Evans and M.M. Telo Da Gama, {\it Molec. Phys.} {\bf 38}
687 (1979).

\bibitem{AbrahamJCP1976}
F.F. Abraham, {\it J. Chem. Phys.} {\bf 64}, 2660 (1976).

\bibitem{Bray}
A.J. Bray, {\it Adv. in Phys.}, {\bf 43}, 357 (1994).

\bibitem{Allen-Cahn}
S.M. Allen and J.W. Cahn, {\it Acta. metall.}, {\bf 27}, 1085 (1979).

\bibitem{footnote2}
Starting from the square gradient functional (\ref{eq:F_GL}) it is
straightforward to show that $1-\rho_b \hat{c}(k)=\beta \rho_b[(\partial^2 f_0/
\partial \rho^2)_{\rho_b} \,+\, Kk^2]$ \cite{Evans79}.

\bibitem{footnote1}
We note that there is some similarity between the structure of Eq.\
(\ref{eq:Spin_dec_eq_2_next}) and the equations of Mode-Coupling Theory -- see
for example E. Zaccarelli, G. Foffi, P. De Gregorio, F. Sciortino, P. Tartaglia
and K.A. Dawson, {\it J. Phys.: Condens. Matter}, {\bf 14}, 2413 (2002) --
used to describe the glass transition.

\bibitem{MaoetalEurPhysJE2001}
Y. Mao, T.C.B. McLeish, P.I.C. Teixeira and D.J. Read, {\it Eur. Phys. J. E},
{\bf 6}, 69 (2001).

\bibitem{Likos}
See e.g.\ C.N. Likos, {\it Phys. Rep.} {\bf 348}, 267 (2001).

\bibitem{AO1}
S. Asakura and F. Oosawa,  {\it J. Chem. Phys.} {\bf 22}, 1255 (1954).

\bibitem{AO2}
S. Asakura and F. Oosawa,  {\it J. Polym. Sci.} {\bf 33}, 183 (1958).

\bibitem{BraderetalMolecPhys2003}
For a recent summary see J.M. Brader, R. Evans and M. Schmidt, {\it Molec.
Phys.}, {\bf 101}, 3349 (2003).

\bibitem{RosenfeldPRL1989}
Y. Rosenfeld, {\it Phys. Rev. Lett.}, {\bf 63}, 980 (1989).

\bibitem{Rosenfeld:Levesque:WeisJCP1990}
Y. Rosenfeld, D. Levesque and J.-J. Weis, {\it J. Chem. Phys.}, {\bf 92},
6818 (1990).

\bibitem{RosenfeldJCP1990}
Y. Rosenfeld, {\it J. Chem. Phys.}, {\bf 93}, 4305 (1990).

\bibitem{Ashcroft:LecknerPhysRev1966}
N.W. Ashcroft and J. Lekner {\it Phys. Rev.}, {\bf 145}, 83 (1966).

\bibitem{AbrahametalJCP1979}
F.F. Abraham, M.R. Mruzic and G.M. Pound, {\it J. Chem. Phys.} {\bf 70}, 2577
(1979).

\bibitem{Lodge:HayesJCP1998}
J.F.M. Lodge and D.M. Heyes, {\it J. Chem. Phys.} {\bf 109}, 7567 (1998).

\bibitem{footnote3}
In Ref.\ \cite{MaoetalEurPhysJE2001} the Flory-Huggins-de Gennes form is used
for $F[C(\rr,t)]$.

\bibitem{EbneretalPRA1976}
C. Ebner, W.F. Saam and D. Stroud, {\it Phys. Rev. A}, {\bf 14}, 2267 (1976).

\bibitem{TDGamma:EvansMolecP1979}
M.M. Telo Da Gama and R. Evans, {\it Molec. Phys.} {\bf 38}
367 (1979).

\end{thebibliography}
\end{document}